\documentclass[%
 reprint,
superscriptaddress,
 amsmath,amssymb,
 aps,
]{revtex4-2}

\usepackage{graphicx}
\usepackage{dcolumn}
\usepackage{bm}
\usepackage[dvipsnames]{xcolor}
\usepackage{hyperref}
\usepackage{ulem}

\begin{document}

\title{Code Comparison for Fast Flavor Instability Simulation}

\author{Sherwood Richers}
\affiliation{Department of Physics, University of California Berkeley, California 94720, USA}
\email{srichers@berkeley.edu}

\author{Huaiyu Duan}
\affiliation{Department of Physics \& Astronomy, University of New Mexico, Albuquerque, New Mexico 87131, USA}

\author{Meng-Ru Wu}

\affiliation{Institute of Physics, Academia Sinica, Taipei 11529, Taiwan}
\affiliation{Institute of Astronomy and Astrophysics, Academia Sinica, Taipei 10617, Taiwan}

\author{Soumya Bhattacharyya}
\affiliation{Institute of Physics, Academia Sinica, Taipei 11529, Taiwan}
\affiliation{Tata Institute of Fundamental Research, Homi Bhabha Road, Mumbai 400005, India}

\author{Masamichi Zaizen}
\affiliation{Faculty of Science and Engineering, Waseda University, Tokyo 169-8555, Japan}
\affiliation{Department of Astronomy, Graduate School of Science, University of Tokyo, Tokyo 113-0033, Japan}

\author{Manu George}
\affiliation{Institute of Physics, Academia Sinica, Taipei 11529, Taiwan}

\author{Chun-Yu Lin}
\affiliation{National Center for High-performance computing, National Applied Research Laboratories, Hsinchu Science Park, Hsinchu City
30076, Taiwan}

\author{Zewei Xiong}
\affiliation{GSI Helmholtzzentrum f\"ur Schwerionenforschung, Planckstra{\ss}e 1, 64291 Darmstadt, Germany}

\date{\today}

\begin{abstract}
The fast flavor instability (FFI) is expected to be ubiquitous in core-collapse supernovae and neutron star mergers. It rapidly shuffles neutrino flavor in a way that could impact the explosion mechanism, neutrino signals, mass outflows, and nucleosynthesis. The variety of initial conditions and simulation methods employed in simulations of the FFI prevent an apples-to-apples comparison of the results. We simulate a standardized test problem using five independent codes and verify that they are all faithfully simulating the underlying quantum kinetic equations under the assumptions of axial symmetry and homogeneity in two directions. We quantify the amount of numerical error in each method and demonstrate that each method is superior in at least one metric of this error. We make the results publicly available to serve as a benchmark.
\end{abstract}

\maketitle


\section{Introduction}
Neutrinos energize the explosions of massive stars and drive outflows from neutron star mergers and proto-neutron stars. In both systems, interactions between neutrinos and outflows determine the elements that form and enrich the universe \cite{Qian:1996xt,muller_HydrodynamicsCorecollapseSupernovae_2020,sarin_EvolutionBinaryNeutron_2020,radice_DynamicsBinaryNeutron_2020}. Furthermore, electron flavor neutrinos and antineutrinos interact more strongly with matter than other flavors due to the large masses of muons and taons, and it is effectively only this flavor of neutrino that is able to convert neutrons to protons and vice versa. The fact that neutrinos can change their flavor in-flight yields a complicated relationship between neutrino flavor and observable properties of supernovae and compact object mergers (see \cite{capozzi_NeutrinoFlavorConversions_2022} for a recent review).

Interactions between neutrinos and other neutrinos are expected to drive rapid and nonlinear evolution of neutrino flavor in these extreme astrophysical environments. A rich variety of flavor transformation phenomena have been found resulting from the mean-field neutrino quantum kinetic equations, including collective flavor transformations \cite{duan_CollectiveNeutrinoOscillations_2010}, the matter-neutrino resonance \cite{malkus_NeutrinoOscillationsBlack_2012}, the neutrino halo effect \cite{cherry_NeutrinoScatteringFlavor_2012}, collisional instability \cite{johns_CollisionalFlavorInstabilities_2021}, and more.
In this work we focus on the fast flavor instability (FFI) \cite{sawyer_MultiangleInstabilityDense_2009,dasgupta_TemporalInstabilityEnables_2015}, another flavor transformation mechanism that has the potential to drive neutrino flavor change in particularly important regions that are inaccessible to other flavor transformation phenomena. In a supernova, this instability is expected to be present above the shock front, beneath the shock front, and in the convecting protoneutron star (\cite{nagakura_WhereWhenWhy_2021} and references therein). Following a neutron star merger, the FFI is expected to be ubiquitous near and inside the resulting accretion disk, precisely where the generation of the universe's heavy elements is thought to occur \cite{wu_FastNeutrinoConversions_2017,george_FastNeutrinoFlavor_2020,Li:2021vqj,Just:2022flt}.

While instability of a distribution can be determined analytically \cite{banerjee_LinearizedFlavorstabilityAnalysis_2011,chakraborty_SelfinducedNeutrinoFlavor_2016,capozzi_FastFlavorConversions_2017,izaguirre_FastPairwiseConversion_2017,abbar_FastNeutrinoFlavor_2018,yi_DispersionRelationFast_2019,chakraborty_ThreeFlavorNeutrino_2020,morinaga_FastNeutrinoFlavor_2021,Dasgupta:2021gfs}, numerical simulations are as of yet required to determine the fate of the distribution after the instability saturates (though see \cite{padilla-gay_NeutrinoFlavorPendulum_2021,bhattacharyya_FastFlavorDepolarization_2021,bhattacharyya_ElaboratingUltimateFate_2022} for analytical work on restricted classes of models). Unfortunately, the spatial and time scales on which the instability operates (sub-cm, sub-nanosecond) are much shorter than the scales of the explosive processes they affect (tens of kilometers and seconds), so direct global simulation of the neutrino quantum kinetics in the full system is, to put it lightly, presently not possible. In order to begin searching for a solution to this conundrum, one can pluck out a small piece of the explosion, i.e., small enough that the neutrino and matter fields look approximately homogeneous, and simulate the instability in that domain only.

To this end, several methods of simulating the FFI have arisen in recent years, each carrying their own set of assumptions and numerical techniques. For simplicity, many calculations impose symmetries in spatial, momentum, or flavor degrees of freedom. Since the neutrino-neutrino interactions driving the FFI do not depend on neutrino energy, it is overwhelmingly common to integrate out the neutrino energy so that the momentum space has at most two direction dimensions. Many early calculations were performed in a beam model, in which all neutrinos are moving in one of two directions \cite{raffelt_NeutrinoFlavorPendulum_2013,dasgupta_FastNeutrinoFlavor_2018,capozzi_CollisionalTriggeringFast_2019}. The neutrino line model, useful for its geometric simplicity, is a initial-value problem that assumes homogeneity and isotropy in one Cartesian direction and allows inhomogeneity and anisotropy in the other (the third spatial dimension is the direction along which the calculation progresses) \cite{duan_FlavorInstabilitiesNeutrino_2015a,abbar_FastNeutrinoFlavor_2019,shalgar_NeutrinoPropagationHinders_2020,padilla-gay_MultiDimensionalSolutionFast_2021}. One can alternatively assume that the neutrino distribution remains axially symmetric around some direction, usually taken to be the radial direction for application to a core-collapse supernova, and impose homogeneity or periodic boundary conditions \cite{dasgupta_FastNeutrinoFlavor_2017,dasgupta_SimpleMethodDiagnosing_2018,johns_FastOscillationsCollisionless_2020,johns_NeutrinoOscillationsSupernovae_2020,martin_DynamicFastFlavor_2020,abbar_SuppressionScatteringInducedFast_2021,bhattacharyya_FastFlavorDepolarization_2021,duan_FlavorIsospinWaves_2021,kato_NeutrinoTransportMonte_2021,martin_FastFlavorOscillations_2021,padilla-gay_FastFlavorConversion_2021,padilla-gay_NeutrinoFlavorPendulum_2021,sasaki_DynamicsFastNeutrino_2021,shalgar_ChangeDirectionPairwise_2021,shalgar_ThreeFlavorRevolution_2021,sigl_SimulationsFastNeutrino_2021,wu_CollectiveFastNeutrino_2021,xiong_StationarySolutionsFast_2021,zaizen_NonlinearEvolutionFast_2021}. Most methods assume two neutrino flavors for simplicity, but there are a growing number of three-flavor simulations of the FFI \cite{richers_ParticleincellSimulationNeutrino_2021,richers_NeutrinoFastFlavor_2021,shalgar_SymmetryBreakingInduced_2021,shalgar_ThreeFlavorRevolution_2021,zaizen_NonlinearEvolutionFast_2021}. There are relatively few methods that account for all of the angular degrees of freedom \cite{richers_ParticleincellSimulationNeutrino_2021,shalgar_DispellingMythDense_2021,shalgar_SymmetryBreakingInduced_2021} (though see similar calculations of the Multi-Azimuthal-Angle Instability \cite{mirizzi_MultiazimuthalangleEffectsSelfinduced_2013,raffelt_AxialSymmetryBreaking_2013,zaizen_ThreeflavorCollectiveNeutrino_2020}), and this has only recently been combined with a treatment of two \cite{bhattacharyya_FastFlavorOscillations_2021} and three \cite{richers_NeutrinoFastFlavor_2021} spatial dimensions.

There are as many choices of initial conditions as there are codes. Given the increasing complexity of the codes, there is a need to understand which aspect of the results are a result of numerical approximations and which are physical results of the evolution equations. Lacking physical data against which to directly validate results, a common approach is to verify that each code is faithfully solving the differential equations via a comparison between codes (e.g., \cite{liebendorfer_SupernovaSimulationsBoltzmann_2005,richers_DetailedComparisonMultidimensional_2017,just_CorecollapseSupernovaSimulations_2018,oconnor_GlobalComparisonCorecollapse_2018,varma_Comparison2DMagnetohydrodynamic_2021}). Such comparisons will be increasingly important in the future as the physics included in simulations becomes more sophisticated (e.g. collisions \cite{richers_NeutrinoQuantumKinetics_2019,capozzi_CollisionalTriggeringFast_2019,johns_CollisionalFlavorInstabilities_2021,shalgar_ChangeDirectionPairwise_2021} and matter inhomogeneities \cite{sigl_SimulationsFastNeutrino_2021}).

In this work we demonstrate good agreement between several codes in the literature on a standardized test problem in one spatial dimension, axial symmetry in momentum space, and two neutrino flavors. In Section~\ref{sec:methods} we summarize the salient features of each simulation method compared in this work. We describe our carefully-defined test problem in Section~\ref{sec:problem} and show the results of the simulations in Section~\ref{sec:results}. Finally, we summarize and provide some discussion in Section~\ref{sec:conclusions}. The numerical data presented here are available at \cite{richers_DatasetCodeComparison_2022}.

\section{Methods}
\label{sec:methods}
In this work we assume the mixing of two neutrino flavors, $e$ and $x$.
The flavor state of a neutrino can be described either in terms of the polarization vector $\mathbf{P} = (P_1, P_2, P_3)$ or the density matrix $\rho$, where the polarization vector components are defined as
\begin{equation}
\begin{aligned}
    P_1 &:= \mathrm{Tr}(\rho \sigma^x) = 2 \mathrm{Re}(\rho_{ex}),\\
    P_2 &:= \mathrm{Tr}(\rho \sigma^y) = -2 \mathrm{Im}(\rho_{ex}),\\
    P_3 &:= \mathrm{Tr}(\rho \sigma^z) = \rho_{ee}-\rho_{xx},\\
\end{aligned}
\label{eq:polarization_vector}
\end{equation}
and $\sigma^i$ are Pauli matrices. We collectively refer to the flavor-coherent components of the polarization vector with the complex quantity $S = P_1 - iP_2$. 

For the sake of a common test problem, we assume the neutrino distributions remain axially symmetric around $\hat{z}$ and are homogeneous along $\hat{x}$ and $\hat{y}$. The direction of a neutrino with velocity $\vec{v}$ is then specified by $u=\hat{v}\cdot\hat{z}$. Under these assumptions, the neutrino distribution evolves according to the quantum kinetic equation
\begin{equation}
(\partial_t + u \partial_z) \rho = -i \left[\mathcal{H},\rho\right]\,\,,
\end{equation}
or equivalently,
\begin{equation}
  (\partial_t + u \partial_z) \mathbf{P} = \boldsymbol{\mathcal{H}}\times \mathbf{P}\,\,.
  \label{eq:QKE}
\end{equation}
We neglect contributions from non-neutrino interaction sources and from the neutrino mass to focus on the pure FFI. Furthermore, we assume $\bar{\rho}=\rho^*$ for antineutrinos (discussed below) so that
\begin{equation}
  \mathcal{H}(z,u) = \int_{-1}^1 du' (1-uu')\left[\mu g(u') - \bar{\mu}\bar{g}(u')\right]\rho(z,u')\,\,.
\end{equation}
In the above expression, $\mu=\sqrt{2}G_F n_\nu$ is the characteristic strength of the Hamiltonian, where $G_F$ is the Fermi coupling constant and $n_\nu$ is the total number density of all neutrino flavors (and similarly for $\bar{\mu}$ for antineutrinos). $g(u)$ and $\bar{g}(u)$ describe the angular distribution of the neutrinos and antineutrinos with normalization $\int du\,g(u) = \int du\,\bar{g}(u) = 1$. $\boldsymbol{\mathcal{H}}$ follows the same vector representation as the density matrix in Eq.~\eqref{eq:polarization_vector}.

Although not all of the codes in this work make all of these assumptions, the initial conditions are carefully constructed such that the simulations are logically equivalent to this form of the equations. In particular, the self-interaction contribution to the Hamiltonian considered here obeys $\bar{\mathcal{H}}=-\mathcal{H}^*$, implying that $\partial_t(\rho-\bar{\rho}^*)=0$. If $\rho=\bar{\rho}^*$ is true in the initial conditions, it remains true throughout the simulation up to numerical error. Axial symmetry is also enforced in the initial conditions for codes that do not assume it.

\subsection{\sc Emu}
{\sc Emu} is a particle-in-cell method for simulating neutrino flavor transformation in a periodic box. The neutrino radiation field is represented by a large number of individual computational particles. Each particle carries with it the number of physical neutrinos $N$ and antineutrinos $\bar{N}$ it represents, the density matrix $\rho_{ab}$ ($\bar{\rho}_{ab}$) common to each (anti-)neutrino in the computational particle, the position $\vec{r}$, and the momentum $\vec{p}$ of each anti/neutrino. Each particle contributes to the number density and number flux vector stored on a background Cartesian grid, which is then used to determine the Hamiltonian for each individual particle. The particles are integrated forward in time using that Hamiltonian and translating at the speed of light. The full PIC implementation uses of a second order shape function and an unsplit forth-order Runge Kutta time integrator. Further details are described in \cite{richers_ParticleincellSimulationNeutrino_2021} and the code is publicly available \cite{willcox_EmuV1_2021}.

\subsection{\sc NuGas}
{\sc NuGas} is a {\sc Python} package that computes collective flavor oscillations of dense neutrino media. The {\sc f2e0d1a} module, which is used in this comparison, implements simple quadrature rules (composite Simpson's rule in this case) for the integration over the neutrino angular distributions and the Lax42 algorithm, a high-precision variant of the two-step Lax-Wendroff method developed by Joshua Martin \cite{martin_NonlinearFlavorDevelopment_2019, martin_DynamicFastFlavor_2020}, for spatial differentiation and temporal integration. The details of the Lax42 algorithm is explained in \cite{duan_FlavorIsospinWaves_2021}. {\sc NuGas} is publicly available through GitHub \cite{nugas}.

\subsection{\sc Cose$\nu$}
The {\sc Cose}$\nu$ code used in \cite{wu_CollectiveFastNeutrino_2021} evolves the components of the density matrix discretized in space and polar angle assuming azimuthal symmetry with two different methods for the advection. 
The {\sc Cose$\nu$-FD} version evaluates advection terms using a fourth-order finite difference method with third-order Kreiss-Oliger artificial dissipation. 
The {\sc Cose$\nu$-FV} version adopts a finite volume method with seventh order WENO reconstruction to evaluate advection terms. 
In both versions, the time evolution is performed with a fourth-order Runge-Kutta scheme. 
The code is publicly available at \url{https://github.com/COSEnu/COSEnu}, 
and detailed description of the code structure, test results, and 
performance is documented in~\cite{George:2022lwg}.

\subsection{Bhattacharyya et al.}
The Bhattacharyya code used in \cite{bhattacharyya_LatetimeBehaviorFast_2020, bhattacharyya_FastFlavorDepolarization_2021, bhattacharyya_FastFlavorOscillations_2021} is written in \textsc{Python}. The main principle behind this numerical setup is to convert a set of coupled nonlinear partial differential equations to a set of coupled nonlinear ordinary differential equations. The code discretizes space ($\vec{r}$) and momentum direction ($\vec{v}$) into equally spaced bins and thus converts the set of coupled nonlinear partial differential equations into a same set of ordinary differential equations (ODE) as a function of time for each $\left(\vec{r}, \vec{v}\right)$ pair. The number of discretized bins are chosen to obtain sufficient accuracy and precision as well as to trigger as many fourier modes as possible especially the unstable ones within limited CPU hours. The total set of ODE's are solved in a finite spatial domain as a function of time using \textsc{Zvode} solver, a variable coefficient differential equation solver in \textsc{SciPy} \cite{virtanen_SciPyFundamentalAlgorithms_2020}, which implements the backward differentiation formula for numerical integration. The code uses the Fast Fourier Transform method implemented in \textsc{scipy.fftpack.diff} package to calculate the gradient term at each spatial location. The differential equation solver adapts the timestep based on target relative and absolute errors. The integrator in this simulation was allowed a relative and absolute error of $10^{-12}$ for $0\leq t \leq 1474\mu^{-1}$ and a relative and absolute error of $10^{-9}$ for $1474\leq t \leq 5000\mu^{-1}$. The change was made to speed up the calculation while maintaining acceptably low errors.

\subsection{Zaizen et al.}
The Zaizen code evolves the Fourier components of the polarization vector discretized in wavenumber and neutrino direction with a fourth-order Runge-Kutta scheme in time. This code adopts a pseudo-spectral method in evaluating the advection term and computes the nonlinear mode-coupling term in the Hamiltonian using the Fast Fourier Transformation implemented in the \textsc{FFTW3} library \footnote{Fastest Fourier Transform in the West, http://www.fftw.org} according to the convolution theorem. To align the simulation setup with others, initial conditions are first built on configuration space and then converted into Fourier space. The spatial Fourier modes are discretized by the inverse of the simulation box size $L$ in this work (by vacuum frequency in a recent application~\cite{zaizen_NonlinearEvolutionFast_2021}). Also, this code adopts the Gauss-Legendre quadrature for the angular integration and arranges the angular distribution on the roots of Legendre polynomials.

\section{Problem Description}
\label{sec:problem}
Here we define a common test problem to simulate based on the neutrino distributions in \cite{martin_DynamicFastFlavor_2020,wu_CollectiveFastNeutrino_2021} and specify initial perturbations with a random spectrum in order to seed the growth of the fast flavor instability. All codes use the same spatial and angular resolution and the same domain size. We do not control the size of the timesteps, as some methods are adaptive and others limit the timestep using a Courant factor.

\subsection{Electron Lepton Number Distribution}
\begin{table}
    \centering
    \begin{tabular}{c|ccc}
        Flavor & $\sqrt{2}G_F n_{\nu_i} / \mu$ & $\sigma$ & A \\ \hline
        $\nu_e$ &  $1$ & 0.6 & 1.33095\\
        $\bar{\nu}_e$ & $0.9$ & 0.53 & 1.50568\\
        $\nu_x$ & 0 & -- & -- \\
    \end{tabular}
    \caption{Parameters for the initial angular distribution of neutrinos as used in Eq.~\eqref{eq:initial_eln}.}
    \label{tab:simulations}
\end{table}

We adopt an electron lepton number (ELN) distribution corresponding to the G3a distribution in \cite{martin_DynamicFastFlavor_2020}. That is, the angular distribution of each neutrino flavor is initially described by
\begin{equation}
\label{eq:initial_eln}
    g(u) = A e^{-(u-1)^2/2\sigma^2}\,\,.
\end{equation}
The normalization constant $A$ is determined by requiring that $\int du\,g(u)=1$. Specifically,
\begin{equation}
    \frac{1}{A} = \sigma\sqrt{\frac{\pi}{2}}  \mathrm{erf}\left(\frac{\sqrt{2}}{\sigma}\right)
\end{equation}
The parameters we use are listed in Table~\ref{tab:simulations}. We choose this ELN distribution because it has already been studied by multiple groups, allowing this verification effort to directly impact those works as well.

\subsection{Perturbations}


The fast flavor instability amplifies unstable modes seeded by perturbations in the initial conditions. We perturb $P_1$ and $P_2$ (or equivalently, $\rho_{ex}$) according to
\begin{equation}
  S(t=0,z) = \sum_{a=-a_\mathrm{max}}^{a_\mathrm{max}} B_a e^{i(k_a z + \phi_a)}
  \label{eq:perturbation}
\end{equation}
where $k_a=2\pi a/L$ with $L$ being the length of the periodic box along $z$. We cut off the spectrum of the perturbations at $a_\mathrm{max}= N_z/20$, where $N_z$ is the number grid cells along $z$, in order to avoid small-scale structure in the perturbations that might induce numerical errors in some methods. This causes the smallest wavelength of perturbations to be resolved by 20 grid cells. The amplitudes of each sinusoid are arbitrarily chosen to be
\begin{equation}
\begin{aligned}
    B_{a=0} = 0
    \qquad\text{and}\qquad 
    B_{a\neq 0}=10^{-7} |a|^{-1}\,\,.
\end{aligned}
\end{equation}
We also choose the phase $\phi_a$ to be uniformily random, sampled independently for each $a$, and not synchronized between the different simulations. The perturbations are isotropic in that $B_a$ and $\phi_a$ are the same for all $u$. Following perturbations to $P_1$ and $P_2$, $P_3$ is adjusted to preserve $|\mathbf{P}|=1$.

\subsection{Simulation Grid}
In order to come as close as possible to the calculations of \cite{martin_DynamicFastFlavor_2020,wu_CollectiveFastNeutrino_2021}, we adopt a simulation box of size $L=10240\mu^{-1}$ spanned by a uniform grid of $N_z=10240$ cells. This choice of simulation domain, together with the above perturbation amplitude and ELN distribution, allow the instability to saturate long before neutrinos are able to wrap around the simulation domain. In addition, we use 200 polar angular bins (or 201 bins in the case of {\sc NuGas}) uniformily-spaced in $u$. In the Zaizen code, angular bins are not uniform but set on the roots of Legendre polynomials. The PIC calculations do not have angular bins, per se, but instead distribute 400 particles around the equatorial direction, which results in approximately 200 polar angles (i.e., 400 particles are needed to represent the single direction $u=0$ in other methods). We assume homogeneity in the $\hat{x}$ and $\hat{y}$ directions, and impose periodic boundary conditions in the $\hat{z}$ direction. We limit the duration of the simulations to $t_\mathrm{max}=5000 \mu^{-1}$ in order to prevent potential consequences of the periodic boundary conditions. This resolution was chosen based on a resolution study using the {\tt NuGas} and \textsc{Cose$\nu$}-FV codes; doubling the spatial resolution caused the polarization vector to be different by at most 0.12 ({\tt NuGas}) or 0.0035 (\textsc{Cose$\nu$}) anywhere on the domain at the end of the simulation. The excellent agreement between methods with different convergence properties (and therefore different amounts of numerical error) suggests that the results are not significantly influenced by the resolution.

\section{Results}
\label{sec:results}

\begin{figure}
    \centering
    \includegraphics[width=\linewidth]{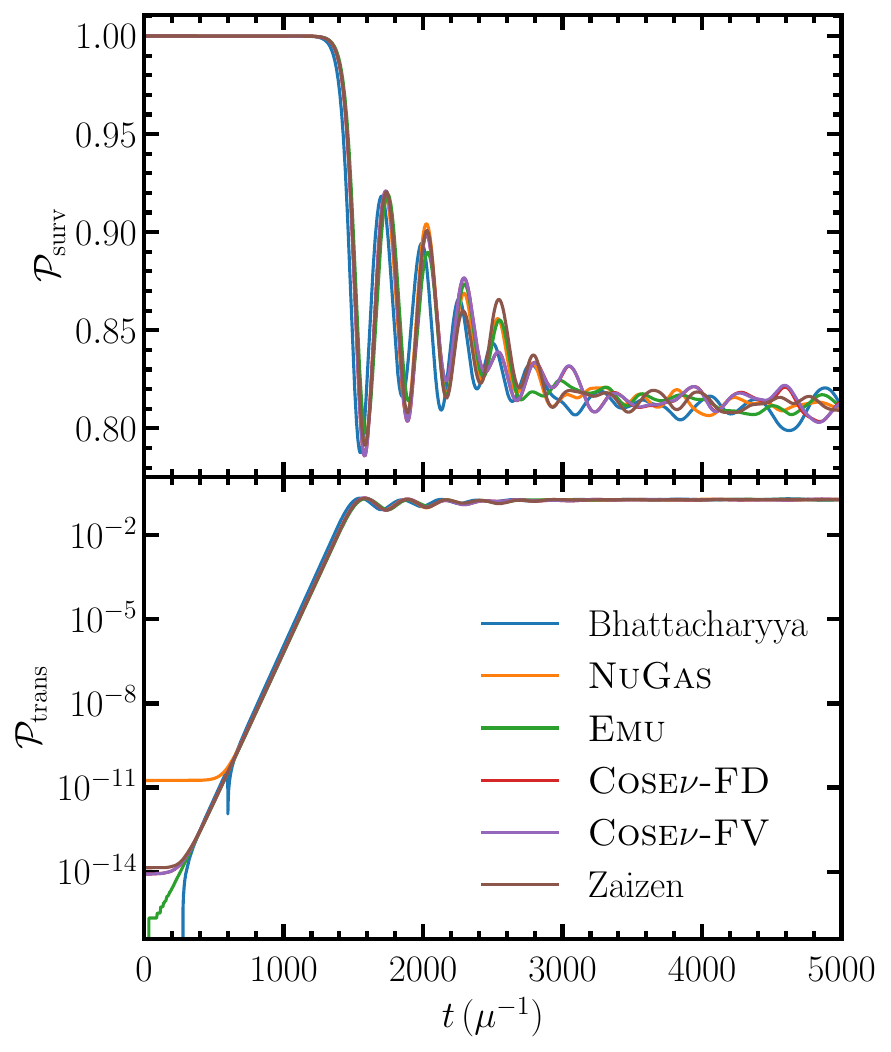}
    \caption{Domain-integrated survival property (top panel) and transition probability (bottom panel) as a function of time. The initial perturbations grow exponentially until the instability saturates at $t\approx1300\mu^{-1}$. All simulations show the same instability growth rate, saturation time, saturation amplitude, and late-time equilibrium.}
    \label{fig:psurv_t}
\end{figure}
We first show good agreement in the average amount of flavor transformation over time. The fraction of neutrinos that remain in the electron flavor state (i.e.~the survival probability) can be expressed as
\begin{equation}
    \mathcal{P}_\mathrm{surv}(t) = \int_{-1}^1 g_{\nu_e}(u) \frac{\langle P_3(t,u) \rangle+1}{2} du,
\end{equation}
where the spatially-averaged polarization vector is
\begin{equation}
    \langle \mathbf{P}(t,u)\rangle = \frac{1}{L}\int_0^{L} \mathbf{P}(t,z,u) dz\,\,.
\end{equation}
We plot the survival probability and the transition probability $\mathcal{P}_\mathrm{trans}=1-\mathcal{P}_\mathrm{surv}$ in Fig.~\ref{fig:psurv_t} with a different color for each method. The results are remarkably similar between simulations, especially given the very different numerical methods and different realizatins of the random perturbations. The survival probability has a value close to 1 during $0<t\lesssim 1300 \mu^{-1}$ as the perturbations grow.  
This can be seen in the bottom panel, which shows a transition probability growing exponentially during that timeframe. The differing floor values of $\mathcal{P}_\mathrm{surv}$ near $t=0$ are a result of differing amounts of floating-point error realized in the different methods, but once the transition probabilities rise above this floor, they line up very closely and grow with an indistinguishable rate. We will discuss numerical error in more detail below.

Once the instability saturates at $t\approx1300 \mu^{-1}$, the survival probability oscillates for a few cycles with approximately the same amplitude and frequency in all of the simulations. The oscillations damp out as the distribution decoheres, and after $t\gtrsim 3000\mu^{-1}$ the survival probability in all of the simulations fluctuate about $\mathcal{P}_\mathrm{surv}\approx 0.82$. By this time, the different simulations are in very different microscopic realizations of the same macroscopic state as a result of the randomized initial conditions.

\begin{figure}
    \centering
    \includegraphics[width=\linewidth]{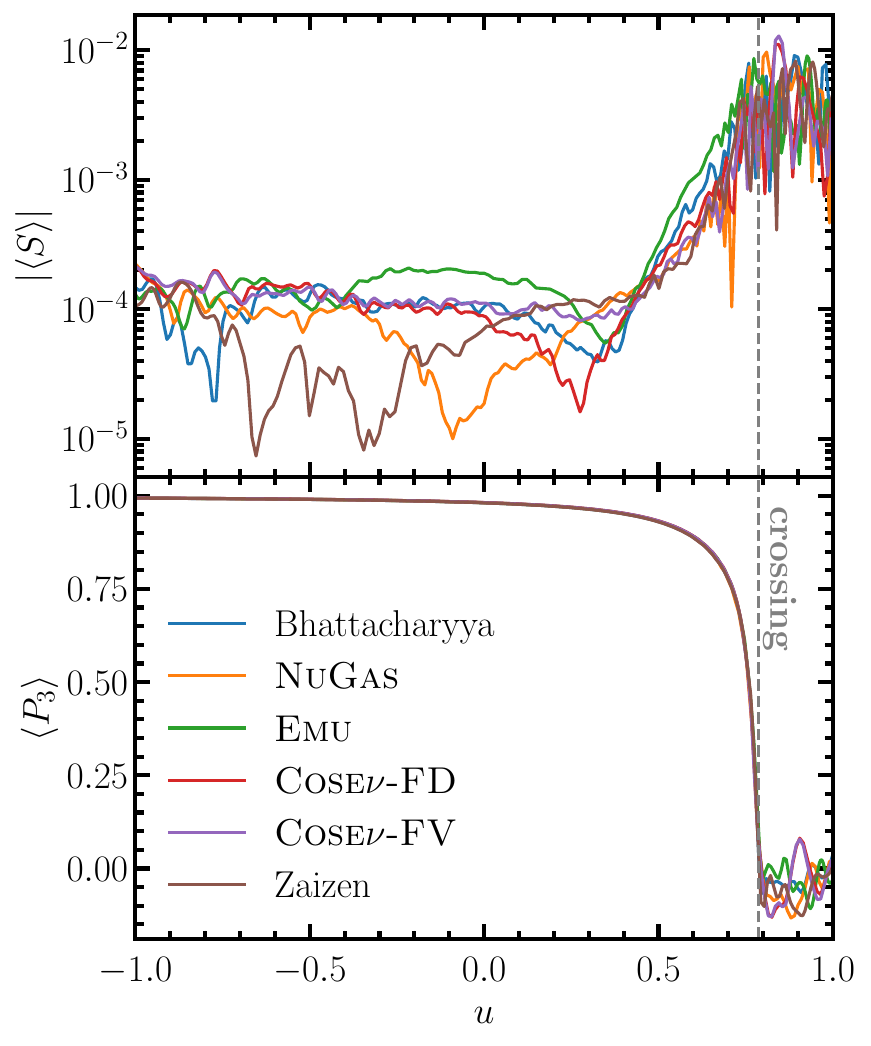}
    \caption{Space-integrated polarization vector components as a function of direction at $t=5000\mu^{-1}$. The vertical dashed line at $u=0.786$ shows the location of the ELN crossing in the initial distribution. All simulations agree on the distribution of neutrino flavor to the left of the crossing and show near complete mixing to the right of the crossing. All simulations agree on the magnitude the flavor off-diagonal components at the end of the simulation.}
    \label{fig:angular_dist}
\end{figure}
Fig.~\ref{fig:angular_dist} shows the spatial average of the flavor vector components at $t=5000\mu^{-1}$. The top panel shows the magnitude of the flavor-coherent (i.e., transverse) components of the polarization vector. The small values indicate that by this point the flavor-coherent components of the polarization vectors at different locations largely cancel each other because of a persisting wave-like pattern in space (see below). The bottom panel shows that the polarization vectors have settled to a well-defined flavor distribution to the left of the crossing (vertical dashed line). To the right of the crossing the neutrinos are fluctuating just below $\langle P_3\rangle=0$, or complete flavor mixing. All of this is in good agreement with \cite{richers_ParticleincellSimulationNeutrino_2021,wu_CollectiveFastNeutrino_2021}.

\begin{figure}
    \centering
    \includegraphics[width=\linewidth]{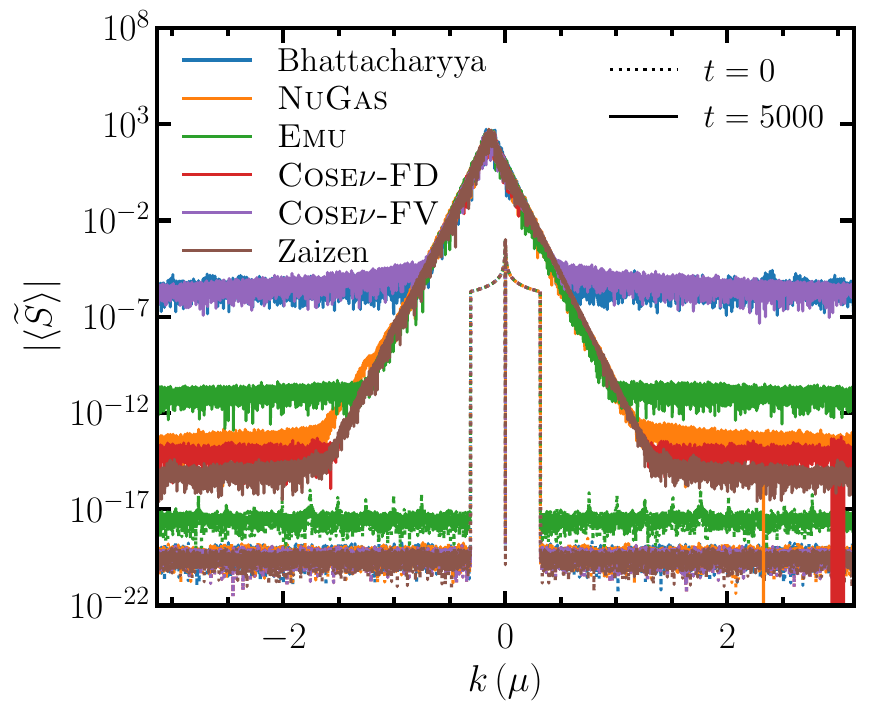}
    \caption{Fourier power spectrum of the flavor coherent off-diagonal component of the neutrino number density $n_{ex}$ at $t=0$ (dotted) and $t=5000$ (solid). All simulations are seeded with the same spectrum of perturbations but with random phases, and all simulations agree on the location of the peak and the slope of the exponential tails at late times. The horizontal bands are a result of numerical error.}
    \label{fig:fft}
\end{figure}
The Fourier spectrum of the distribution also shows excellent agreement throughout the simulation. We compute the number-weighted direction-averaged power spectrum given by
\begin{equation}
    \langle \widetilde{S}(t,k)\rangle=\int_{-1}^1 g(u)du \int_0^L e^{-i k z}  S(t,z,u) dz\,\,.
\end{equation}
The power spectrum of the initial perturbation common to all simulations and described in Eq.~\eqref{eq:perturbation} is apparent in the dotted curves in Fig.~\ref{fig:fft}. The solid curves show the power spectrum at the end of the simulation ($t=5000\mu^{-1}$). By this point, the unstable modes have already grown and saturated. Even at this late time, all methods show excellent agreement. The horizontal bands in both the initial and final spectra are a result of numerical errors, and the Zaizen code shows the smallest error in this metric. As suggested in \cite{richers_ParticleincellSimulationNeutrino_2021} (for different choices of neutrino distribution), the resulting power spectrum is static, with exponential tails away from the peak. The peak of the equilibrium spectrum is not at $k=0$, reflecting the presence of long-lived coherent wave-like pattern in the spatial distribution of the polarization vectors as demonstrated by \cite{duan_FlavorIsospinWaves_2021} and observed in the TwoThirds simulation of \cite{richers_NeutrinoFastFlavor_2021}. This coherent wave structure is not apparent in the upper panel of Fig.~\ref{fig:angular_dist} because the data there are spatially integrated over many periods, yielding a number close to 0. Although the exponential tails seem to be a robust feature of these simulations, we still lack a satisfactory explanation for them.

\begin{figure}
    \centering
    \includegraphics[width=\linewidth]{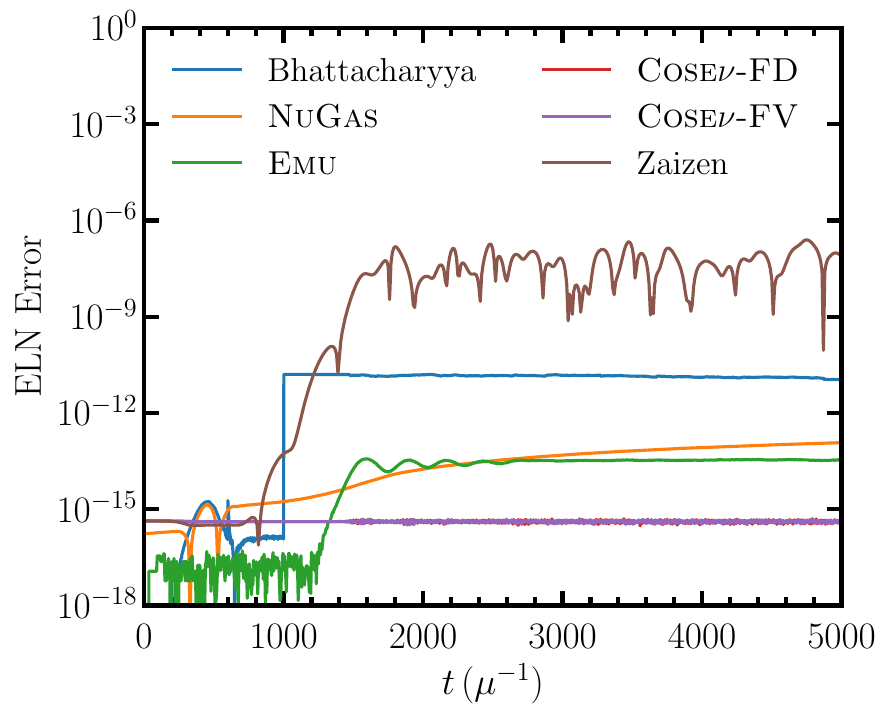}
    \caption{Deviation of the domain-integrated electron lepton number from its initial value, as defined in Eq.~\eqref{eq:eln_error}. The SU(2) symmetry of the Hamiltonian guarantees this remain at zero, so nonzero values reflect numerical error. All codes exhibit excellent ELN conservation to better than one part in $10^6$.}
    \label{fig:ndens_error}
\end{figure}
The SU(2) symmetry of the neutrino self-interaction Hamiltonian should preserve the net electron lepton number (ELN) of the neutrino distribution. As a test of the quality of the numerical scheme, we show the violation of this constraint in Fig.~\ref{fig:ndens_error}. Specifically, we define
\begin{align}
    \mathrm{ELN\,Error} &= 
    \frac{1}{n_{\nu_e} + \bar{n}_{\nu_e}}
    \Bigg|\int_{-1}^{1}[n_{\nu_e}g_{\nu_e}(u)-\bar{n}_{\nu_e}\bar{g}_{\nu_e}(u)]
    \nonumber\\
    &\quad\times\left(\frac{1-\langle P_3(t,u)\rangle}{2}\right)d u\Bigg|
    \,\,.
    \label{eq:eln_error}
\end{align}
where $n_{\nu_e}$ and $\bar{n}_{\nu_e}$ are the initial electron neutrino and antineutrino number densities listed in Table~\ref{tab:simulations}. This quantity probes the self-interaction term in Eq.~\eqref{eq:QKE} more strongly than the advection term. In all cases, the error remains smaller than one part in $10^6$. The error grows most significantly during the linear growth phase, even growing exponentially with the perturbation amplitude in some codes. After saturation, the error continues to grow sub-linearly with time at a rate that is not visible on this plot for most codes. Although {\sc Emu} (green) has the lowest error for the first $1200\mu^{-1}$ time units, the error then quickly grows above both of the \textsc{Cose$\nu$} codes, which maintain remarkably low ELN error throughout the duration of the simulation.

We found that in general using an angular integration method during post-processing that is inconsistent with that used to model the evolution equations shows significantly and artificially large ELN errors. For instance, artificially large error is reported if during the simulation angular integrals are performed with Simpson's rule but in post-processing the integrals are performed with the pyramid rule. Similarly, artificially large errors can be reported if the code does not restrict $P_3=\bar{P}_3$ but assumes so in post-processing. The continuous and finite-difference evolution equations based on only the neutrino-neutrino potential (in combination with our choice of initial conditions) both guarantee that $P_3=\bar{P}_3$, but only up to floating-point precision, allowing finite precision errors that violate this guarantee to accumulate in time. Finally, errors can be introduced by assuming that $P_3^2=1-P_1^2-P_2^2$, which is numerically true only up to floating point precision. Each code made a particular combination of choices, and we found that errors are minimized when the post-processing methods make the same assumptions as the underlying code.

\begin{figure}
    \centering
    \includegraphics[width=\linewidth]{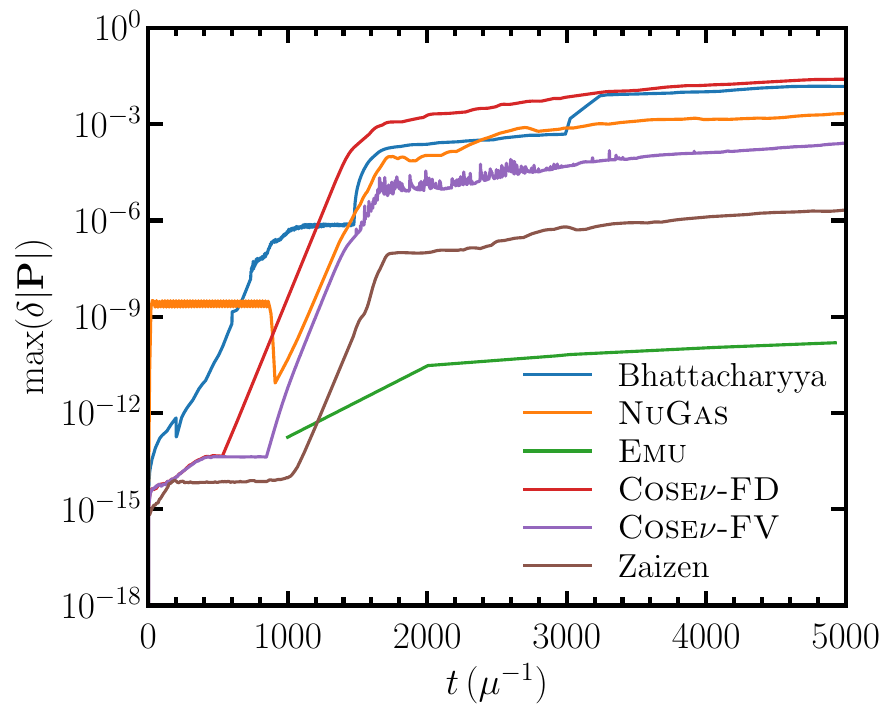}
    \caption{Maximum deviation of the flavor vector length from unity. The Hermitian nature of the Hamiltonian guarantees that this remain at zero. Nonzero values reflect numerical error, especially in the advection terms. There are few Emu data points resulting from infrequent output of the storage-intensive particle data.}
    \label{fig:length_error}
\end{figure}
The Hermitian nature of the Hamiltonian also guarantees that the length of the physical polarization vector does not change length. Since all polarization vectors start with unit magnitude, the deviation from this at a given space-direction bin is
\begin{equation}
    \delta|\mathbf{P}| = \left|\sqrt{P_1^2+P_2^2+P_3^2}-1\right|\,\,.
\end{equation}
In Fig.~\ref{fig:length_error} we show the evolution of the maximum of this quantity over all direction bins and spatial grid cells as a probe of numerical error. Once again, all codes show excellent results. The vector length error grows exponentially during the linear growth phase, following the exponential growth of the perturbations. After saturation, the error grows super-linearly in time. These errors are most strongly affected by the advection term in Eq.~\eqref{eq:QKE}. 
Here the Bhattacharyya, \textsc{NuGas} and \textsc{Cose}$\nu$ codes, which evaluate the advection terms with discretized spatial grids, yield relatively larger errors compared to
the Zaizen code and \textsc{Emu}. In particular, the {\sc Emu} results show the lowest error because the particle nature of the code eliminates advection errors, and the errors shown are a result of only the self-interaction term.

\section{Conclusions}
\label{sec:conclusions}
All of the codes presented robustly predict the instability growth rate, saturation amplitude, angular distribution, and post-saturation Fourier spectrum of the neutrino distribution. As suggested by \cite{wu_CollectiveFastNeutrino_2021,richers_ParticleincellSimulationNeutrino_2021,bhattacharyya_ElaboratingUltimateFate_2022,bhattacharyya_FastFlavorDepolarization_2021}, the neutrinos within the ELN crossing (i.e., directions dominated by the less abundant species) undergo near complete flavor mixing, while neutrinos outside the ELN crossing only exhibit partial transformation. As demonstrated by \cite{duan_FlavorIsospinWaves_2021} and observed in the TwoThirds simulation of \cite{richers_NeutrinoFastFlavor_2021}, the post-saturation distribution maintains modes that do not decay away. The exponential tail of the post-saturation spectrum observed in \cite{richers_NeutrinoFastFlavor_2021} is robustly produced by all codes.

Each simulation exhibits small numerical errors, though no method is consistently better or worse than others in all metrics. As one might expect, the Lagrangian method in \textsc{Emu} yields small advection errors, and the Zaizen code, which operates fully in the Fourier domain, has the smallest errors in Fourier space. The \textsc{Cose}$\nu$ code, which actively enforces ELN conservation, maintains low ELN error. We naturally find that timestep size and integration method impact the magnitude of the errors, though the adaptive nature of some codes precluded a uniform timestep choice. We also find that it is particularly important to use a post-processing integration method that is consistent with that used to evolve the distribution in order to accurately report errors.

Since global simulations of neutrino quantum kinetics are currently not possible, practitioners of local simulations are forced to pick their poison when it comes to initial conditions and boundary conditions. One approach, as we have done here, is to perturb the entire domain and assume periodic boundary conditions. This choice reflects an expectation that the background distribution is homogeneous on the scale of the simulation domain and that perturbations in adjacent domains look like those in the simulated domain. Another approach is to provide a single local perturbation and end the simulation before the boundary conditions come into play, thereby ensuring that any results are not a consequence of the choice of artificial boundary conditions. Both are unrealistic, because a supernova is not infinitely periodic and nature is unlikely to ensure that perturbations in different locations never interact with each other. \cite{wu_CollectiveFastNeutrino_2021} takes the middle ground and provide a local perturbation, but simulate with periodic boundary conditions for more than a domain traversal time. The results look similar to but distinct from the same simulations with random perturbations. This work lends confidence to the robustness of simulation results given artificial initial and boundary conditions, setting the stage for work toward more realistic simulations.

\begin{acknowledgments}
We are grateful for the New Directions in Neutrino Flavor Evolution in Astrophysical Systems at the Institute for Nuclear Theory in September 2021, which sparked discussions that led to this paper. SR is supported by the NSF Astronomy and Astrophysics Postdoctoral Fellowship under Grant No. 2001760.
HD is supported by the U.S.\ Department of Energy, Office of Science, Office of Nuclear Physics under Award Number DE-SC0017803.
MZ is supported by the Japan Society for Promotion of Science (JSPS) Grant-in-Aid for JSPS Fellows (Grants No. 20J13631) from the Ministry of Education, Culture, Sports, Science, and Technology (MEXT) in Japan. The numerical computations were carried out on Cray XC50 at the Center for Computational Astrophysics, National Astronomical Observatory of Japan, Pride and Flock computing clusters in the Department of Theoretical Physics at TIFR Mumbai.
MRW, MG and SB acknowledge supports from the Ministry of Science and Technology, Taiwan under Grant No.~110-2112-M-001 -050, and
the Academia Sinica under Project No.~AS-CDA-109-M11.
MRW also acknowledge supports from the Physics Division, National Center for Theoretical Sciences, Taiwan.
MRW and MG appreciate the computing resources provided by the Academia Sinica Grid-computing Center. SB also acknowledges the support from Department of Theoretical Physics, TIFR Mumbai.
CYL thank the National Center for High-performance Computing (NCHC) for providing computational and storage resources.
ZX acknowledge supports from the European Research Council (ERC) under the European Union's Horizon 2020 research and innovation programme (ERC Advanced Grant KILONOVA No.~885281).
\end{acknowledgments}

\bibliography{references}

\end{document}